\documentclass[twoside]{dis07}
\usepackage[latin1]{inputenc}
\usepackage[dvips]{graphicx,epsfig,color}
\usepackage{wrapfig,rotating}
\usepackage{amssymb,amsmath,array}

\begin{document}
\title{The Curvature of $F_2^p(x, Q^2)$ as a Probe of Perturbative QCD
Evolutions in the small--$x$ Region}

%***********************************************************************
% AUTHORS INFORMATION AREA
%***********************************************************************
\author{Cristian Pisano
%
% Optional short acknowledgment: remove next line if non-needed
%\thanks{This is an optional funding source acknowledgment.}
%
% DO NOT MODIFY THE FOLLOWING '\vspace' ARGUMENT
\vspace{.3cm}\\
%
% Addresses and institutions (remove "1- " in case of a single institution)
Department of Physics and Astronomy, Vrije Universiteit Amsterdam, \\
De Boelelaan 1081, 1081 HV Amsterdam - The Netherlands 
%1- School of First Author - Dept of First Author \\
%Address of First Author's school - Country of First Author's
%school
%
% Remove the next three lines in case of a single institution
%\vspace{.1cm}\\
%2- School of Second Author - Dept of Second Author \\
%Address of Second Author's school - Country of Second Author's school\\
}
%***********************************************************************
% END OF AUTHORS INFORMATION AREA
%***********************************************************************

\maketitle

%%%%%%%%%%%%%%%%%%%%%%%%%%%%%%%%%%%%%%%%%%%%%%%%%%%%%%%%%%%%%%%%%%%%%%%%%
\begin{abstract}
Perturbative NLO and NNLO QCD evolutions of parton distributions
are studied, in particular in the (very) small-$x$ region, where they are 
in very good agreement with all recent precision measurements of 
$F_2^p(x,Q^2)$.  These predictions turn out to be also rather insensitive
to the specific choice of the factorization scheme ($\overline{\rm MS}$ or
DIS).  A characteristic feature of perturbative QCD evolutions is a 
{\em{positive}} curvature of $F_2^p$ which increases as $x$ decreases.
This perturbatively stable prediction provides a sensitive test of the 
range of validity of perturbative QCD.
\end{abstract}
%%%%%%%%%%%%%%%%%%%%%%%%%%%%%%%%%%%%%%%%%%%%%%%%%%%%%%%%%%%%%%%%%%%%%%%%

The curvature of DIS structure functions like $F_2^p(x,Q^2)$, i.e., its
second derivative with respect to the photon's virtuality $Q^2$ at
fixed values of $x$, plays a decisive role in probing the range of 
validity of perturbative QCD evolutions of parton distributions in the 
small-$x$ region.  This has been observed recently 
\cite{url,ref1,ref2,ref2b} and
it was demonstrated that NLO($\overline{\rm MS}$) evolutions
imply a {\em positive} curvature which increases as $x$ decreases.
Such  rather unique predictions provide a check of the range of validity
of perturbative QCD evolutions.  However, the curvature is a rather 
subtle mathematical quantity which a priori may sensitively depend on
the theoretical (non)perturbative assumptions made for calculating it.
Our main purpose is to study the dependence and
stability of the predicted curvature with respect to a different choice
of the factorization scheme (DIS versus $\overline{\rm MS}$) and to 
the perturbative order of the evolutions by extending the common NLO
(2-loop) evolution \cite{ref2} to the next-to-next-to-leading 3-loop
order (NNLO) \cite{ref2b}.

\begin{table}
\normalsize
\renewcommand{\arraystretch}{1.8} 
%\parbox{17cm}{}
\scriptsize
\centering 
\begin{tabular}{|c||c|c|c|c||c|c|c|c|}
\hline
& \multicolumn{4}{|c||}{NNLO($\overline{\rm MS}$)}  & 
\multicolumn{4}{|c|}{NLO($\overline{\rm MS}$)} \\   
%\multicolumn{4}{|c|}{NLO(DIS)}\\
%******************************************************
\hline
& $u_v$ & $d_v$ & $\bar{q}$ & $g$ &
  $u_v$ & $d_v$ & $\bar{q}$ & $g$ \\
%  $u_v$ & $d_v$ & $\bar{q}$ & $g$\\
%******************************************************
\hline
N & 0.2503 & 3.6204 & 0.1196 & 2.1961 &
    0.4302 & 0.3959 & 0.0546 & 2.3780 \\
%    0.6885 & 0.4476 & 0.0702 & 2.3445\\

a & 0.2518 & 0.9249 & -0.1490 & -0.0121 &
    0.2859 & 0.5375 & -0.2178 & -0.0121 \\
%    0.3319 & 0.5215 & -0.1960 & -0.0121\\ 

b & 3.6287 & 6.7111 & 3.7281 & 6.5144 &
    3.5503 & 5.7967 & 3.3107 & 5.6392 \\
%    2.6511 & 2.290 & 5.5480 & 6.8581\\

c & 4.7636 & 6.7231 & 0.6210 & 2.0917 &
    1.1120 & 22.495 & 5.3095 & 0.8792  \\
%    -1.6163 & 10.398 & 3.7277 & 1.8732\\

d & 24.180 & -24.238 & -1.1350 & -3.0894 &
    15.611 & -52.702 & -5.9049 & -1.7714 \\
%    15.197 & -16.466 & -4.7067 & -2.4302\\

e & 9.0492 & 30.106 & --- & --- &
    4.2409 & 69.763 & --- & ---  \\
%&    -7.6056 & 5.6364 & --- & ---\\
\hline
%****************************************************** 
$\chi^2/{\rm dof}$ & \multicolumn{4}{|c||}{0.989} & 
                     \multicolumn{4}{|c|}{0.993} \\ 
%                     \multicolumn{4}{|c|}{0.992}\\
\hline
%******************************************************
$\alpha_s(M_Z^2)$ & \multicolumn{4}{|c||}{0.112} & 
                      \multicolumn{4}{|c|}{0.114} \\ 
%&                      \multicolumn{4}{|c|}{0.114}\\
\hline
%******************************************************
\end{tabular}
\caption{Parameter values of the NLO and NNLO QCD fits
with the parameters of the input distributions referring to (1)
and (2).}
\end{table}
The valence $q_v=u_v,\, d_v$ and sea
$w=\bar{q},\, g$ distributions underlying $F_2^p(x, Q^2)$ 
are parametrized at an input scale
$Q_0^2=1.5$ GeV$^2$ as follows:
%Eqs.(12)+(13)
\begin{eqnarray}
x\,q_v(x,Q_0^2) & = & N_{q_v}x^{a_{q_v}}(1-x)^{b_{q_v}}
  (1+c_{q_v}\sqrt{x}+d_{q_v}x + e_{q_v}x^{1.5})\\
x\,w(x,Q_0^2) & = & N_w x^{a_w}(1-x)^{b_w}(1+c_w\sqrt{x}+d_w x)
\end{eqnarray}
and without loss of generality the strange sea is taken to be 
$s=\bar{s}=0.5\, \bar{q}$. Notice that we 
do not consider sea breaking effects ($\bar{u}\neq\bar{d},\,\, s\neq
\bar{s}$) since the data used, and thus our analysis, are not
sensitive to such corrections. The normalizations $N_{u_v}$ and $N_{d_v}$
are fixed by $\int_0^1 u_v dx = 2$ and $\int_0^1 d_v dx=1$,
respectively, and $N_g$ is fixed via $\int_0^1 x(\Sigma +g)dx=1$.
We have performed all $Q^2$-evolutions in Mellin $n$-moment space and used
the QCD-PEGASUS program \cite{ref10} for the NNLO evolutions. For 
definiteness we work in the fixed flavor factorization scheme, 
rather than in the variable (massless quark) scheme
since the results for $F_2^p$ and its curvature remain essentially 
unchanged \cite{ref2}.

We have somewhat extended the set of DIS data used in \cite{ref2}
in order to determine the remaining parameters at larger values
of $x$ and of the valence distributions. 
The following data sets
have been used:  the small-$x$ \cite{ref14} and large-$x$ 
\cite{ref15} H1 $F_2^p$ data; the fixed target BCDMS data 
\cite{ref16} for $F_2^p$ and $F_2^n$ using $Q^2\geq 20$ GeV$^2$
and $W^2=Q^2(\frac{1}{x}-1)+m_p^2\geq 10$ GeV$^2$ cuts, and the 
proton and deuteron NMC data \cite{ref17} for $Q^2\geq 4$ GeV$^2$
and $W^2\geq 10$ GeV$^2$.  This amounts to a total of 740 data 
points.  The required overall normalization factors of the data
are 0.98 for BCDMS and 1.0 for NMC.
The resulting parameters of the NLO($\overline{\rm MS}$) and 
NNLO($\overline{\rm MS}$) fits are summarized in Table~1.  

The quantitative difference between the NLO($\overline{\rm MS}$)
and NLO(DIS) results turns out to be rather small \cite{ref2b}.  Therefore
we do not consider any further the DIS scheme in NNLO. 
%The relevant small-$x$ predictions are compared with
%the H1 data \cite{ref14} in Fig.~1, which are also consistent with
%the ZEUS data \cite{ref18} with partly lower statistics.  
The present more detailed NLO($\overline{\rm MS}$) analysis corresponds
to $\chi^2/{\rm dof}=715.3/720$ and the results are comparable
to our previous ones \cite{ref2}. 
Our new NLO(DIS) and NNLO(3-loop) fits are also very similar, 
corresponding to $\chi^2/{\rm dof}=714.2/720$ and $712.0/720$, respectively.
%%%%%%%%%%%%%%%%%%%%%%%%%%%%%%%%%%%%%%%%%%%%%%%%%%%%%%%%%%%%%%%%%%%%%%%%%%
\begin{wrapfigure}{r}{0.5\columnwidth}
\centerline{\includegraphics[width=0.5\columnwidth]
{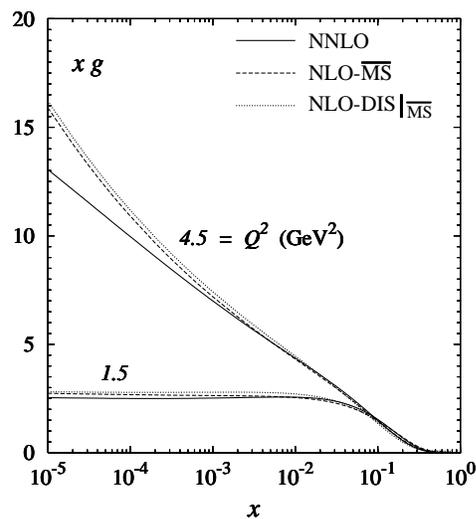}}
\caption{The gluon distributions at the input scale $Q_0^2=1.5$ 
GeV$^2$ and at $Q^2=4.5$ GeV$^2$.}
\end{wrapfigure}
%%%%%%%%%%%%%%%%%%%%%%%%%%%%%%%%%%%%%%%%%%%%%%%%%%%%%%%%%%%%%%%%%%%%%%%%%%%
%%%%%%%%%%%%%%%%%%%%%%%%%%%%%%%%%%%%%%%%%%%%%%%%%%%%%%%%%%%%%%%%%%%%%%%%%%%
%, although they fall slightly
%below the common NLO($\overline{\rm MS}$) predictions at smaller
%values of $Q^2$. 
It should be emphasized that the perturbatively
stable QCD predictions are in perfect agreement with all recent
high-statistics measurements of the $Q^2$-dependence of 
$F_2^p(x,Q^2)$ in the (very) small-$x$ region.  Therefore
additional model assumptions concerning further resummations of
subleading small-$x$ logarithms (see, for example, \cite{ref19})
are not required \cite{ref7,ref9}.

Figure~1  shows our gluon input distributions in (1) and Table~1 
as obtained in our three
different fits, as well as their evolved shapes at $Q^2=4.5$ GeV$^2$
in particular in the small-$x$ region. In order to allow for a
consistent comparison in the $\overline{\rm MS}$ scheme, our
NLO(DIS) results have been transformed to the $\overline{\rm MS}$
factorization scheme.  Note, however, that the
gluon distribution in the DIS scheme is very similar to the one
obtained in NLO($\overline{\rm MS}$) shown in Fig.~1 which holds in
particular in the small-$x$ region.  This agreement becomes even
better for increasing values of $Q^2$.  This agreement is similar
for the sea distributions in the small-$x$ region.
Only for 
$x$ \raisebox{-0.1cm}{$\stackrel{>}{\sim}$} 0.1 the NLO(DIS) sea
density becomes sizeably smaller than the NLO($\overline{\rm MS}$)
one. The NLO results are rather similar but
distinctively different from the NNLO ones in the very small-$x$
region at $Q^2>Q_0^2$.  In particular the strong increase of the
gluon distribution $xg(x,Q^2)$ as $x\to 0$ at NLO is somewhat
tamed by NNLO 3-loop effects.

Turning now to the curvature of $F_2^p$ we first present in Fig.~2
our results for $F_2^p(x,Q^2)$ at $x=10^{-4}$, together with a 
global fit MRST01 NLO result \cite{ref20}, as a function of \cite{ref1}
%Eq.(14)
\begin{equation}
q = \log_{10}\left(1+\frac{Q^2}{0.5\,\,{\rm GeV}^2}\right)\, \, .
\label{eq:q}
\end{equation}
This variable has the advantage that most measurements lie along
a straight line \cite{ref1} as indicated by the dotted line in 
Fig.~2.  All our three NLO and NNLO fits give almost the same
results which are also very similar \cite{ref2} to the global
CTEQ6M NLO fit \cite{ref21}.  In contrast to all other fits shown in
Fig.~2, only the MRST01 parametrization results in a sizeable
curvature for $F_2^p$.  More explicitly the curvature
can be directly extracted from
%Eq.(15)
\begin{equation}
F_2^p(x,Q^2) = a_0(x) + a_1(x)q + a_2(x)q^2\,\, .
\label{eq:f2}
\end{equation}
The curvature $a_2(x)=\frac{1}{2}\,\partial_q^2\,  F_2^p(x,Q^2)$ is
evaluated by fitting this expression to the predictions for
$F_2^p(x,Q^2)$ at fixed values of $x$ to a (kinematically) given
interval of $q$.  In Figure~3 we present $a_2(x)$ which results 
from experimentally selected $q$-intervals \cite{ref1,ref2,ref2b}:
%Eq.(16)
\begin{eqnarray}
0.7 \leq q \leq 1.4\quad\quad & {\rm for} & \quad\quad
    2\times 10^{-4} < x < 10^{-2}\nonumber\\
0.7 \leq q \leq 1.2\quad\quad & {\rm for} & \quad\quad
    5\times 10^{-5} < x \leq 2\times 10^{-4}\, .
\label{eq:interval}
\end{eqnarray}
%%%%%%%%%%%%%%%%%%%%%%%%%%%%%%%%%%%%%%%%%%%%%%%%%%%%%%%%%%%%%%%%%%%%%%%%%%%
%%%%%%%%%%%%%%%%%%%%%%%%%%%%%%%%%%%%%%%%%%%%%%%%%%%%%%%%%%%%%%%%%%%%%%%%%%%
\begin{wrapfigure}{r}{0.5\columnwidth}
\centerline{\includegraphics[width=0.5\columnwidth]
{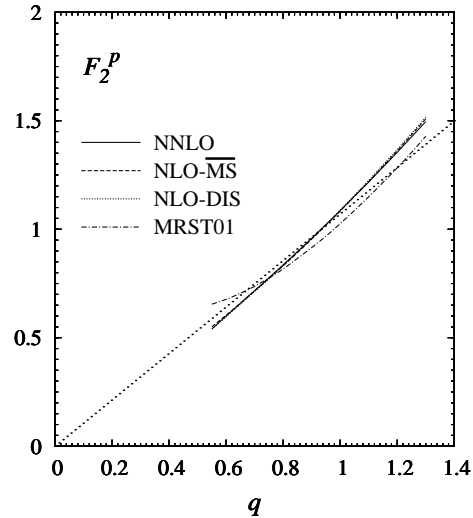}}
\caption{Predictions for $F_2^p(x,Q^2)$ at $x=10^{-4}$ plotted
versus $q$ defined in (\ref{eq:q}).}
\end{wrapfigure}
%%%%%%%%%%%%%%%%%%%%%%%%%%%%%%%%%%%%%%%%%%%%%%%%%%%%%%%%%%%%%%%%%%%%%%%%%%%
It should be noticed that the average value of $q$ decreases with
decreasing $x$ due to the kinematically more restricted $Q^2$ range
accessible experimentally. (We deliberately do not show the results at
the smallest available $x=5\times 10^{-5}$ where the $q$-interval is
too small, $0.6\leq q\leq 0.8$, for fixing $a_2(x)$ in (\ref{eq:f2}) uniquely
and where moreover present measurements are not yet sufficiently
accurate).
% For comparison we also show in 
%Fig.~5b the curvature $a_2(x)$ for an $x$-independent fixed $q$-interval
%%Eq.(17)
%\begin{equation}
%0.6\leq q \leq 1.4 \quad\quad\quad 
%      (1.5\leq Q^2\leq 12\,\, {\rm GeV}^2)\, \, .
%\end{equation}  
Apart from the rather large values of $a_2(x)$ specific \cite{ref2,ref2b}
for the MRST01 fit, our NLO and NNLO results agree well with the 
experimental curvatures as calculated and presented in \cite{ref1}
using the H1 data \cite{ref14}.  Our predictions do 
{\em{not}} sensitively depend on the factorization scheme
chosen ($\overline{\rm MS}$ or DIS) and are, moreover, perturbative
{\em{stable}} with the NNLO 3-loop results lying typically
below the NLO ones, i.e.\ closer to present data \cite{ref2b}.  
It should be
emphasized that the perturbative stable evolutions always result
in a {\em{posi\-tive}} curvature which {\em{increases}}
as $x$ decreases.  Such unique predictions provide a sensitive
test of the range of validity of perturbative QCD!  
This feature is supported by the data shown in Fig.~3. Future
analyses of present precision measurements in this very small-$x$
region (typically
$10^{-5}$ \raisebox{-0.1cm}{$\stackrel{<}{\sim}$} $x$
\raisebox{-0.1cm}{$\stackrel{<}{\sim}$} $10^{-3}$) should 
provide additional tests of the theoretical predictions concerning
the range of validity of perturbative QCD evolutions.
%%%%%%%%%%%%%%%%%%%%%%%%%%%%%%%%%%%%%%%%%%%%%%%%%%%%%%%%%%%%%%%%%%%%%%%%%%
\begin{wrapfigure}{r}{0.5\columnwidth}
\centerline{\includegraphics[width=0.5\columnwidth]
{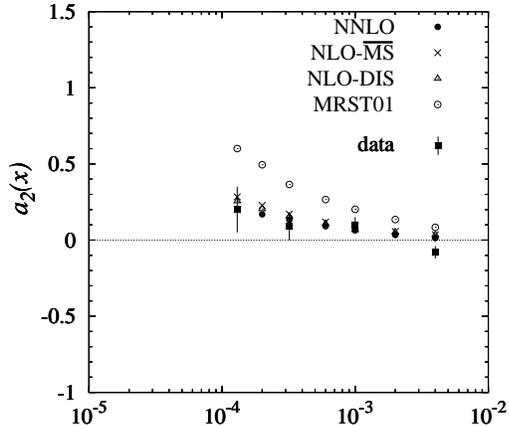}}
\caption{The curvature $a_2(x)$ as defined in (\ref{eq:f2}) for  the
variable $q$-intervals in (\ref{eq:interval}).}
\end{wrapfigure}
%%%%%%%%%%%%%%%%%%%%%%%%%%%%%%%%%%%%%%%%%%%%%%%%%%%%%%%%%%%%%%%%%%%%%%%%%%%
%%%%%%%%%%%%%%%%%%%%%%%%%%%%%%%%%%%%%%%%%%%%%%%%%%%%%%%%%%%%%%%%%%%%%%%%%%%

To conclude, perturbative NLO and NNLO QCD evolutions of parton distributions
in the (very) small-$x$ region are fully compatible with all 
recent high-statistics measurements of the $Q^2$-dependence of
$F_2^p(x,Q^2)$ in that region.  The results are perturbatively
stable and, furthermore, are rather insensitive to the 
factorization scheme chosen ($\overline{\rm MS}$ or DIS).
Therefore additional model assumptions concerning further
resummations of subleading small-$x$ logarithms are not required.
A characte\-ristic feature of perturbative QCD evolutions is a
{\em{positive}} curvature $a_2(x)$ which {\em{increases}}
as $x$ decreases  (cf.~Fig.~3). This rather unique and perturbatively
stable prediction plays a decisive role in probing the range of
validity of perturbative QCD evolutions.  Although present data
are indicative for such a behavior, they are statistically 
insignificant for $x<10^{-4}$.  Future analyses of present
precision measurements in the very small-$x$ region should provide
a sensitive test of the range of validity of perturbative QCD and
further information concerning the detailed shapes of the gluon
and sea distributions as well.

% ****************************************************************************
% BIBLIOGRAPHY AREA
% ****************************************************************************

\begin{footnotesize}
% IF YOU DO NOT USE BIBTEX, USE THE FOLLOWING SAMPLE SCHEME FOR THE REFERENCES
% ----------------------------------------------------------------------------

% ----------------------------------------------------------------------------

% IF YOU USE BIBTEX,
% - DELETE THE TEXT BETWEEN THE TWO ABOVE DASHED LINES
% - UNCOMMENT THE NEXT TWO LINES AND REPLACE 'Name_Of_Your_BibFile'

%\bibliographystyle{unsrt}
%\bibliography{Name_Of_Your_BibFile}
% example of Name_Of_Your_BibFile.bib
% @Article{Turcato:2006ch,
%      author    = "Turcato, M.",
%  collaboration = "ZEUS and H1",
%      title     = "Lepton flavour violation and charmonium physics at HERA",
%      journal   = "Nucl. Phys. Proc. Suppl.",
%      volume    = "162",
%      year      = "2006", 
%      pages     = "283-287",
%      SLACcitation  = "%%CITATION = NUPHZ,162,283;%%"
% }
% 
% @Unpublished{Gogitidze:2007du,
%      author    = "Gogitidze, N.",
%  collaboration = "H1", 
%      title     = "Prompt photons and particle momentum distributions at
%                   HERA", 
%      year      = "2007",
%      note    = "hep-ex/0701033",
%      SLACcitation  = "%%CITATION = HEP-EX 0701033;%%"
% }

\end{footnotesize}

% ****************************************************************************
% END OF BIBLIOGRAPHY AREA
% ****************************************************************************

\end{document}